\documentclass{article}

\usepackage{amsmath} 
\usepackage{physics} 

\usepackage{xcolor}
\usepackage{graphicx}
\usepackage{subfigure}

\usepackage{booktabs} 
\usepackage{array} 
\usepackage{multirow} 
\usepackage{multicol} 
\usepackage{adjustbox} 
\usepackage{makecell} 
\usepackage{longtable} 
\usepackage{caption} 
\usepackage{algorithm}
\usepackage{algorithmicx}
\usepackage{algpseudocode}

\usepackage{cite}
\makeatletter
\def\@cite#1#2{\textsuperscript{[{#1\if@tempswa , #2\fi}]}} 
\makeatother

\usepackage{amssymb} 
\makeatletter
\newcommand{\Rmnum}[1]{\expandafter\@slowromancap\romannumeral #1@} 
\makeatother

\usepackage[normalem]{ulem} 

\usepackage{cancel} 

\begin{document}
\title{Stochastic Resolution of Identity for Correlation Energy Prediction via Doubles Connected Moments Expansion}
\author{Chongxiao Zhao\textsuperscript{1,2,a)} and Wenjie Dou\textsuperscript{1,2,b),*}}
\date{}
\maketitle 

\begin{center}
    1.Department of Chemistry, School of Science, Westlake University, Hangzhou, Zhejiang 310024, China \\
    2.Institute of Natural Sciences, Westlake Institute for Advanced Study, Hangzhou, Zhejiang 310024, China \\
    \textsuperscript{a)}Email: zhaochongxiao@westlake.edu.cn \\
    \textsuperscript{b)}Authors to whom correspondence should be addressed: douwenjie@westlake.edu.cn \\
\end{center}

\begin{abstract}
The recently developed Doubles Connected Moments (DCM) expansion offers a tractable approach for computing correlation energy, exhibiting an noniterative $O(N^6)$ scaling with system size $N$.
Benchmark calculations on a set of molecules demonstrate that the DCM can outperform CCSD in terms of accuracy.
To further enhance its efficiency, we present a stochastic variant of DCM by introducing a stochastic resolution-of-identity (sRI) technique, which decomposes the essential four-index intermediates.
The resulting sRI-DCM scheme only involves one $O(N^6)$ step, while all other steps do not exceed $O(N^4)$ at each recursion, and reliably reproduces the results of conventional DCM.
Our sRI-DCM achieves an overall experimental scaling of $O(N^{4.46})$ for series hydrogen dimer chains, demonstrating that it is attractive and practical for large systems containing hundreds of electrons.
\end{abstract}

\section*{1. INTRODUCTION}

\hspace{1em} 
The accurate determination of correlation energy, defined as the difference between the exact energy and the approximate energy from HF wavefunction, constitutes a central problem in quantum chemistry.\cite{szabo2012modern}
Traditional post‑HF methods, such as Møller–Plesset perturbation theory (MPn),\cite{moller1934note,bartlett1978many,krishnan1978approximate} configuration interaction (CI),\cite{slater1929theory,bender1969studies} and coupled‑cluster (CC) theory,\cite{coester1958bound,coester1960short} systematically improve the wavefunction mainly by truncating excitations to a certain level, thereby balancing accuracy with computational cost\cite{blinder2018mathematical}
These approaches have achieved remarkable success, often reaching chemical accuracy for a wide range of molecular systems.
However, these traditional methods face significant difficulties in treating strongly correlated systems.
Moreover, some high-level post‑HF methods renders them prohibitively expensive for larger molecules, underscoring the need for alternative computational strategies.

The $t$-expansion\cite{horn1984t} offers an alternative to calculate the correlation energy with moments.
This method is inherently non-perturbative and can capture strong correlation effects if the trial wavefunction $\ket{\Phi}$ has some overlap with the exact ground state.
Introduce an auxiliary function
\begin{eqnarray}
    \begin{aligned}
        \langle E(t) \rangle &= \bra{\Phi} H \ket{\Phi} = \frac{\bra{\Phi} H e^{-tH} \ket{\Phi}}{\bra{\Phi} e^{-tH} \ket{\Phi}} \\
        &= \sum^{\infty}_{m=0}{\frac{(-t)^m}{m!} I_{m+1}} \label{eqn:t-expansion}
    \end{aligned}
\end{eqnarray}
The so-called connected moments $I_m$ (or cumulants) satisfy the following relationship with moments $\mu_m$.
\begin{gather}
    \mu_m = \bra{\Phi} H^m \ket{\Phi} \\
    I_{m+1} = \mu_{m+1} - \sum_{n=0}^{m-1} C^n_m I_{n+1} \mu_{m-n} \label{eqn:cumulant_moment}
\end{gather}
Then the ground state energy $E_0$ is given as
\begin{equation}
    E_0 = \lim_{t \rightarrow \infty}{E(t)}
\end{equation}

In practice, the infinite power series in eq \ref{eqn:t-expansion} needs to be truncated to a certain order.
Among all the possible solutions\cite{stubbins1988methods}, the connected moments expansion (CMX)\cite{cioslowski1987connected,knowles1987validity} expresses the energy with the lowest $2n-1$ connected moments as
\begin{equation}
    E_{\mathrm{CMX(n)}} = I_1 - 
    \begin{bmatrix}
        I_2 & I_3 & \cdots & I_n
    \end{bmatrix}
    \begin{bmatrix}
        I_3     & I_4     & \cdots & I_{n+1} \\
        I_4     & I_5     & \cdots & I_{n+2} \\
        \vdots  & \vdots  & \ddots & \vdots \\
        I_{n+1} & I_{n+2} & \cdots & I_{2n-1} \\
    \end{bmatrix}^{-1}
    \begin{bmatrix}
        I_2 \\
        I_3 \\
        \vdots \\
        I_n \\
    \end{bmatrix}
\end{equation}
Higher-order connected moments are related to lower-order ones via eq \ref{eqn:cumulant_moment}.

As early as 1987, Cioslowski \textit{et al.} tested CMX(2), which recovers more than 50\% of MP3 energy.\cite{cioslowski1987connected_2}
As the CMX method accurately predicts the correlation energy as a noniterative moment-based scheme, however, the computational scaling increases sharply with the order.
For example, $I_4$ contains single, double, triple and quadruple substitutions and gives an $O(N^7)$ scaling, comparable to MP4.
This steep computational scaling ultimately limits this moment-based methods\cite{ai2025quantum,zhuravlev2016cumulant,zhuravlev2020cumulant,zhuravlev2022cumulant} to relatively low orders, constraining their achievable accuracy.

In the DCM formulation reported by Ganoe and Head-Gordon\cite{ganoe2023doubles}, the method operates solely within the double excitation subspace of the Hilbert space.
By constructing four-index intermediates and establishing relationships between connected moments of different orders, the computational cost can be substantially reduced to $O(N^6)$.
DCM achieves smooth convergence within approximately $N$=11-14, with accuracy surpassing that of CCSD with proper reference.
Besides, this moment-based method can handle strong correlated systems, which traditional single-reference methods like CCSD fail to address.

Inspired by the high performance of DCM, we incorporate the stochastic resolution of identity (sRI) technique to further reduce its computational and storage costs.
The sRI is a stochastic formulation built upon the deterministic RI framework\cite{feyereisen1993use,eichkorn1995auxiliary,bernholdt1996large}.
This strategy introduces a set of stochastic orbitals to decouple the high-rank tensors, primarily the four-index electron repulsion integrals (ERIs), into lower-rank integrals, which allows flexible scaling reduction and avoids steep memory need.
So far, the sRI technique has been successfully used in many electronic structrue methods, such as MP2,\cite{neuhauser2013expeditious,ge2014guided,takeshita2017stochastic} 
CC2,\cite{zhao2024stochastic_gs,zhao2024stochastic_ex,zhao2025stochastic,zhao2025stochasticresolutionidentitycc2} 
DFT and TDDFT,\cite{baer2013self,neuhauser2014communication,gao2015sublinear,neuhauser2016stochastic,bradbury2023deterministic,fabian2025compact} 
GF2,\cite{neuhauser2017stochastic,takeshita2019stochastic,dou2019stochastic,dou2020range,mejia2023stochastic,mejia2024convergence} and etc.\cite{rabani2015time,vlvcek2016spontaneous,lee2020stochastic}

In this work, we integrate the sRI approximation to the DCM formulation, abbreviated as sRI-DCM.
This incorporation enables flexible decomposition of intermediate steps and reduces scaling by more than one order of magnitude.
Our sRI-DCM is capable of treating systems with hundreds of electrons, and its accuracy is systematically tunable via the number of stochastic orbitals.
This methods extends the applicational scope of DCM and enables the determination of correlation energy for larger systems.

\section*{2. THEORY}

\hspace{1em} 
The standard notation used in this work is summarized in Table \ref{tab:notation}.
The parameters listed in the final column ($N_{AO}$, $N_{aux}$, $N_{ao}$, $N_{mo}$, $N_{occ}$, and $N_{vir}$) all scales linearly with the system size $N$.
\begin{table}[htbp]
    \caption{Summary of notations in the following equations.} 
    \begin{adjustbox}{center} 
        \begin{tabular}{lcc} 
            \toprule[0.3pt]
            \specialrule{0em}{0.3pt}{0.3pt} 
            \midrule
            \makebox[0.3\textwidth][l]{Items} & \makebox[0.3\textwidth][c]{Functions or indices} & \makebox[0.15\textwidth][c]{Total number} \\ 
            \midrule
            AO Gaussian basis functions & $\chi_{\alpha}(r_1)$, $\chi_{\beta}(r_1)$, $\chi_{\gamma}(r_1)$, $\chi_{\delta}(r_1)$, $\cdots$ & $N_{AO}$ \\
            Auxiliary basis functions   & $P$, $Q$, $R$, $S$, $\cdots$ & $N_{aux}$ \\
            General AOs         & $\alpha$, $\beta$, $\gamma$, $\delta$, $\cdots$ & $N_{ao}$ \\
            General MOs         & $p$, $q$, $r$, $s$, $\cdots$ & $N_{mo}$ \\
            Occupied MOs                & $i$, $j$, $k$, $l$, $\cdots$ & $N_{occ}$ \\
            Unoccupied MOs              & $a$, $b$, $c$, $d$, $\cdots$ & $N_{vir}$ \\
            \midrule
            \specialrule{0em}{0.3pt}{0.3pt}
            \bottomrule[0.3pt]
        \end{tabular}
    \end{adjustbox}
    \label{tab:notation}
\end{table}

\subsection*{2.1. DCM formulation}

\hspace{1em} 
In the DCM formulation, a connected moment of the Hamiltonian—for instance, the second one—is calculated using the moments
\begin{equation}
    I_2 = \langle H^2 \rangle - \langle H \rangle^2
\end{equation}
By approximately inserting the identity $1=P+D$, the equation becomes
\begin{gather}
    P = \ket{\Phi_0} \bra{\Phi_0} \\
    D = \frac{1}{4} \sum_{ijab} \ket{\Phi^{ab}_{ij}} \bra{\Phi^{ab}_{ij}} \\
    I_2 = \langle H (P+D) H \rangle - \langle H P H \rangle = \langle H D H \rangle = \langle H^2\rangle_c
\end{gather}
Here $P$ is constructed from the HF reference $\ket{\Phi_0}$ and $D$ is the doubles subspace of the orthogonal Hilbert space.
Similarly, higher-order connected moments can be expressed in this form.
Their structures are governed by the four-index intermediates $(ij||ab)_n$ as
\begin{align}
    I_{2n-1} = \langle \underbrace{H D \cdots H}_{n H} D \underbrace{H \cdots D H}_{(n-1) H} \rangle = \frac{1}{4} \sum_{ijab}{(ij||ab)_{n} (ij||ab)_{n-1}} \\
    I_{2n} = \langle \underbrace{H D \cdots H}_{n H} D \underbrace{H \cdots D H}_{n H} \rangle = \frac{1}{4} \sum_{ijab}{(ij||ab)_{n} (ij||ab)_{n}}
\end{align}
These intermediates satisfy the following recursion relation
\begin{eqnarray}
    \begin{aligned}
        (ij||ab)_{n+1} &= \bra{0} H \ket{2} \bra{2} H \ket{2}_{n} = \frac{1}{4} \sum_{klcd}{ (kl || cd)_{n}\bra{\Phi^{cd}_{kl}} H \ket{\Phi^{ab}_{ij}}} \\
        &= \frac{1}{2} \sum_{cd}{\langle cd || ab \rangle (ij || cd)_n} + \frac{1}{2} \sum_{kl}{\langle ij || kl \rangle (kl || ab)_n} \\
        &- \Delta^{ab}_{ij} (ij || ab)_n + P(ij) P(ab) \sum_{kc}{\langle kb || ic \rangle (kj || ca)_{n}}
        \label{eqn:recursion}
    \end{aligned}
\end{eqnarray}
Here $0$ denotes $\Phi_0$ while $2$ denotes $\Phi_{\mu_2}$, and
\begin{gather}
    \Delta^{ab}_{ij} = \epsilon_i + \epsilon_j - \epsilon_a - \epsilon_b \\
    P(pq) f(p, q) = f(p, q) - f(q, p)
\end{gather}

The primary determining steps for the $O(N^6)$ scaling of DCM are the contractions of two four-index tensors, the antisymmetrized two-center integrals $\langle pq||rs \rangle$ and intermediates $(ij||ab)_n$, in eq \ref{eqn:recursion}.
By efficiently decoupling these tensors using sRI, it is possible to reduce this scaling while also lowering storage demands.

\subsection*{2.2. RI and sRI formulations}

\hspace{1em} 
In deterministic RI formulations, the four-index ERI is approximated using a combination of three-index $(\alpha\beta|P)$ and two-index $V_{PQ}$ integrals expanded in an auxiliary basis $\{P\}$.
\begin{equation}
    (\alpha \beta|\gamma \delta) \approx \sum_{PR} {(\alpha \beta|P) V^{-1}_{PR} (R|\gamma \delta)} \\
\end{equation}
Introducing the three-rank tensor $B^Q_{\alpha \beta}$, the four-index RI-ERI takes the form
\begin{equation}
    B^Q_{\alpha \beta} = \sum_P {(\alpha \beta|P) V^{-1/2}_{PQ}}
\end{equation}
\begin{equation}
    (\alpha \beta|\gamma \delta) \approx \sum_Q {\left[ \sum_P {(\alpha \beta|P) V^{-1/2}_{PQ}} \right] \left[ \sum_R {V^{-1/2}_{QR} (R|\gamma \delta)} \right]} = \sum_Q {B^Q_{\alpha \beta} B^Q_{\gamma \delta}} \label{eqn:RI-ERI}
\end{equation}
The construction of $B^Q_{\alpha \beta}$ costs $O(N_{aux}^2 N_{ao}^2)$ and the final step scales as $O(N_{aux} N_{ao}^4)$.
Since both $N_{aux}$ and $N_{ao}$ grow linearly with system size, the overall computational scaling for RI-ERI is $O(N^5)$, with a storage requirement of $O(N^3)$.

sRI is a variant of RI that further introduces a set of $N_s$ stochastic orbitals \{$\theta^{\xi}$\}, $\xi=1,2,\cdots,N_s$.
Each stochastic orbital $\theta^{\xi}$ is a column array of length $N_{aux}$, with all its element randomly generated to be 1 or -1.
Thus, as $N_s$ increases, these $\theta^{\xi}$ satisfy the following equation
\begin{equation}
    \left\langle \theta \otimes \theta \right\rangle_\xi =\frac{1}{N_s} \sum_{\xi = 1}^{N_s} {\theta ^\xi \otimes \left( \theta ^\xi \right)^T} \approx I \label{eqn:sRI_I}
\end{equation}
This approximate identity can then be inserted into eq \ref{eqn:RI-ERI} to transform one auxilary basis index $Q$ into a stochastic orbital index $\xi$
\begin{eqnarray}
    \begin{aligned}
        (\alpha \beta|\gamma \delta) &= \left\langle \sum_P {\left[(\alpha \beta|P) \sum_Q {\left( V^{-1/2}_{PQ} \theta_Q^{\xi} \right)} \right]} \sum_R {\left[ (R|\gamma \delta) \sum_S {\left( V^{-1/2}_{SR} \theta_S^{\xi} \right)} \right]} \right\rangle_\xi \\
        &= \left\langle R^{\xi}_{\alpha \beta} R^{\xi}_{\gamma \delta} \right\rangle_\xi = \frac{1}{N_s} \sum_{\xi = 1}^{N_s} {R^{\xi}_{\alpha \beta} R^{\xi} _{\gamma \delta}}
    \end{aligned}
\end{eqnarray}
\begin{equation}
    R^{\xi}_{\alpha \beta} = \sum_P {(\alpha \beta|P) \sum_Q {\left( V^{-1/2}_{PQ} \theta_Q^{\xi} \right)}}
\end{equation}
The construction of $R^{\xi}_{\alpha \beta}$ needs two steps, which respectively scale as $O(N_s N_{aux}^2)$ and $O(N_s N_{aux} N_{ao}^2)$.
And the evaluation of four-index sRI-ERI costs $O(N_s N_{ao}^4)$.

Unlike conventional RI, sRI employs a size-independent parameter $N_s$.
Consequently, the storage requirement for the intermediate tensor $R^{\xi}_{\alpha \beta}$ scales as $O(N_s N_{ao}^2)$, which is effectively $O(N^2)$.
In practice, this reduction by an order of magnitude from the $O(N^3)$ scaling of $B^Q_{\alpha \beta}$ enables more flexible and computationally efficient tensor contractions.
The primary error of sRI originates from eq \ref{eqn:sRI_I} and can be systematically controlled by adjusting $N_s$, which will be discussed in detail later.

\subsection*{2.3. sRI-DCM}

\hspace{1em} 
To align with Ganoe \textit{et al.}’s equations, we retain the spin orbital form in the following derivation.
This allows our code to handle both restricted and unrestricted spin cases.
As the time-determining steps are the contractions of four-index tensors, we mainly focus on the decomposition of these high-order tensors. 

First, the antisymmetrized two-center integrals can be constructed with RI or sRI
\begin{eqnarray}
    \begin{aligned}
        \langle pq||rs \rangle &= \langle pq|rs \rangle - \langle pq|sr \rangle \\
        &= \sum_Q{(B^Q_{pr} B^Q_{qs} - B^Q_{ps} B^Q_{qr})} \\
        &= \langle R^{\xi}_{pr} R^{\xi}_{qs} - R^{\xi}_{ps} R^{\xi}_{qr} \rangle_{\xi}
    \end{aligned}
\end{eqnarray}

Apart from the decomposition of four-index ERIs, the four-index intermediates $(ij||ab)_n$ are also needed to be decoupled.
Since these intermediates do not necessarily possess the positive definiteness property like two-electron integrals, additional effort is required to decompose them.
To this end, we present two decomposition schemes.

The first scheme is to adopt eigendecomposition.
This begins by rearranging $(ij||ab)_n$ from a four-index to a two-index form, expressed in terms of the composite indices ($ia$) and ($jb$).
\begin{eqnarray}
    \begin{aligned}
        (ij||ab)_n &\triangleq A_{ia,jb} = VEV^T \\
        &=
        \begin{pmatrix}
            V^{(-)} \\
            V^{(+)}
        \end{pmatrix}^T
        \left[
        \begin{pmatrix}
            0 & 0 \\
            0 & E^{(+)}
        \end{pmatrix}
        +
        \begin{pmatrix}
            E^{(-)} & 0 \\
            0 & 0
        \end{pmatrix}
        \right]
        \begin{pmatrix}
            V^{(-)} \\
            V^{(+)}
        \end{pmatrix} \\
        &= T_{ia,jb}^{(+)} - T_{ia,jb}^{(-)} = T_{ia,jb}^{(+),\frac{1}{2}} T_{ia,jb}^{(+),\frac{1}{2}} - T_{ia,jb}^{(-),\frac{1}{2}} T_{ia,jb}^{(-),\frac{1}{2}}
    \end{aligned}
\end{eqnarray}
The eigendeompostion itself scales as $O(N^6)$.
Since $A_{ia,jb}$ may contain both positive and negative eigenvalues, we partition them into two positive semidefinite blocks and handle each one separately.
Below we take the construction of positive eigenvalues contribution as an example.
\begin{equation}
    T_{ia,jb}^{(+),\frac{1}{2}} = (V E^{(+), \frac{1}{4}}) (V E^{(+), \frac{1}{4}})^T
\end{equation}

Then transform one index pair of $o$-$v$ into sRI space.
\begin{eqnarray}
    \begin{aligned}
        (ij||ab)_{n} &\approx \sum_{kc}{T_{ia,kc}^{(+),\frac{1}{2}} T_{kc,jb}^{(+),\frac{1}{2}}} - \sum_{kc}{T_{ia,kc}^{(-),\frac{1}{2}} T_{kc,jb}^{(-),\frac{1}{2}}} \\
        &= \sum_{kck'c'}{T_{ia,kc}^{(+),\frac{1}{2}} \left( \left\langle \theta \otimes \theta \right\rangle_\xi \right)_{kc,k'c'} T_{k'c',jb}^{(+),\frac{1}{2}}} - \sum_{kck'c'}{T_{ia,kc}^{(-),\frac{1}{2}} \left( \left\langle \theta \otimes \theta \right\rangle_\xi \right)_{kc,k'c'} T_{k'c',jb}^{(-),\frac{1}{2}}} \\
        &= \left\langle (\sum_{kc}{T_{ia,kc}^{(+),\frac{1}{2}} \theta^\xi_{kc}}) (\sum_{k'c'}{\theta^\xi_{k'c'} T_{k'c',jb}^{(+),\frac{1}{2}}}) - (\sum_{kc}{T_{ia,kc}^{(-),\frac{1}{2}} \theta^\xi_{kc}}) (\sum_{k'c'}{\theta^\xi_{k'c'} T_{k'c',jb}^{(-),\frac{1}{2}}}) \right\rangle_\xi \\
        &= \left\langle T_{ia}^{(+),\xi} T_{jb}^{(+),\xi} - T_{ia}^{(-),\xi} T_{jb}^{(-),\xi} \right\rangle_\xi
    \end{aligned}
\end{eqnarray}
Such a subtle decomposition enables us to handle each tensor contraction involving the $(ij||ab)_{n}$ by replacing it with the two-rank tensors $T_{ia}^{(+),\xi}$ and $T_{ia}^{(-),\xi}$.
The final expression resembles the sRI formulation of four-index ERIs, with matching computational cost and storage scaling.
The difference is that sRI-ERI replaces only one auxiliary basis index with an sRI index, while here we replace a pair of occupied and virtual orbital indices.
This shows that sRI is not limited to decoupling four-index ERIs.
In fact, sRI has the potential to decompose any positive semidefinite four-index tensor into two-rank tensors, which is an attractive feature.

The second scheme is Cholesky decomposition\cite{beebe1977simplifications,epifanovsky2013general,pedersen2024versatility}.
Conventional Cholesky decomposition also scales as $O(N^6)$.
However, we try this technique in the hope that if the targeted intermediates are sparse, the scaling can be reduced to below $O(N^4)$.
Since $A_{ia,jb}$ may not be positive semidefinite, which is a requirement for Cholesky decomposition, we decompose it into two semidefinite components by adding and subtracting a scaled identity matrix $I$.
Specifically, only the first component then requires Cholesky decomposition and the index $J$ is proportional to $N^2$.
\begin{eqnarray}
    \begin{aligned}
        (ij||ab)_n \triangleq A_{ia,jb} &= (A_{ia,jb} + c * I_{ia,jb}) - c * I_{ia,jb} \\
        &= \sum_J{L^J_{ia} L^J_{jb}} - c * I_{ia,jb}
    \end{aligned}
\end{eqnarray}
Here the factor $c$ is chosen as the the inverse of the lowest negative eigenvalue of $A_{ia,jb}$, or $c = 0$ if all eigenvalues are non‑negative.
The calculation of $c$ costs $O(N^4)$.
We do not set a fixed $c$ for all calculations, since as we found in our previous sRI implementations\cite{zhao2025stochastic,zhao2025stochasticresolutionidentitycc2}, such treatment sometimes introduce significant stochastic noise when sRI is later applied to the two semi-definite components.

Subsequently, the sRI approximated identity is inserted into the first component, while the second component is constructed directly using stochastic orbitals.
\begin{eqnarray}
    \begin{aligned}
        (ij||ab)_{n} &\approx \sum_{JJ'}{L^J_{ia} \left( \left\langle \theta \otimes \theta \right\rangle_\xi \right)_{J,J'} L^{J'}_{jb}} - c \left( \left\langle \theta \otimes \theta \right\rangle_\xi \right)_{ia,jb} \\
        &= \left\langle (\sum_J{L^J_{ia} \theta^\xi_J}) (\sum_{J'}{\theta^\xi_{J'} L^{J'}_{jb}}) - (\sqrt{c} \theta^\xi_{ia}) (\sqrt{c} \theta^\xi_{jb}) \right\rangle_\xi \\
        &= \left\langle M_{ia}^{\xi} M_{jb}^{\xi} - N_{ia}^{\xi} N_{jb}^{\xi} \right\rangle_\xi
    \end{aligned}
\end{eqnarray}
The final expression appears similar to that of the first scheme.
Except for the Cholesky decomposition, the scaling of all other steps is also reduced to $O(N^4)$.
Unfortunately, within our test, the sparse density of $A_{ia,jb}$ may be quite high, implying that performing the Cholesky decomposition still incurs a cost of $O(N^6)$.
Finally, the observed scaling of the second scheme is similar to that of the first scheme, while the second scheme shows larger errors.
Therefore, in the subsequent sections, we still use the eigendecompostion and we provide some data in the Appendix for the second scheme.

With the sRI approximation, the recurrence relation in eq \ref{eqn:recursion} now becomes
\begin{eqnarray}
    \begin{aligned}
        (ij||ab)_{n+1} &= \frac{1}{2} \langle \sum_{cd}{(R^{\xi'}_{ca} R^{\xi'}_{db} - R^{\xi'}_{cb} R^{\xi'}_{da})(T_{ic}^{(+),\xi} T_{jd}^{(+),\xi} - T_{ic}^{(-),\xi} T_{jd}^{(-),\xi})} \rangle_{\xi \xi'} \\
        &+ \frac{1}{2} \langle \sum_{kl}{(R^{\xi'}_{ki} R^{\xi'}_{lj} - R^{\xi'}_{kj} R^{\xi'}_{li})(T_{ka}^{(+),\xi} T_{lb}^{(+),\xi} - T_{ka}^{(-),\xi} T_{lb}^{(-),\xi})} \rangle_{\xi \xi'} \\
        &- \Delta^{ab}_{ij} (ij || ab)_n \\
        &+ P(ij) P(ab) \langle \sum_{cd}{(R^{\xi'}_{kj} R^{\xi'}_{bc} - R^{\xi'}_{kc} R^{\xi'}_{bj})(T_{ic}^{(+),\xi} T_{ka}^{(+),\xi} - T_{ic}^{(-),\xi} T_{ka}^{(-),\xi})} \rangle_{\xi \xi'}
    \end{aligned}
\end{eqnarray}
Here, we use two set of stochastic orbitals, labeled $\xi$ and $\xi'$, because the two four-index tensors are independent.
Since all sRI tensors $R_{pq}^{\xi'}$ and $T_{ia}^{(\pm),\xi}$ are two-rank tensors and each index appears twice at most in any single multiplication, the contraction of these two-rank tensors scales as $O(N^3)$ at most.
However, the final assembly of $(ij||ab)_{n}$ requires $O(N^4)$ operations.
Moreover, the eigendeompostion step scales as $O(N^6)$.
These two steps prevent our sRI-DCM formulation from reaching a better $O(N^3)$ scaling.
Nevertheless, this scheme remains valuable because it reduces the number of expensive $O(N^6)$ steps and we can observe scaling reduction in the results.
As for storage, the two-rank sRI tensors add extra $O(N^2)$ memory cost, but this is negligible compared to eliminating the $O(N^4)$ storage needed for $\langle pq||rs \rangle$.

\section*{3. RESULTS AND DISCUSSION}

\hspace{1em} 
Using RI and sRI techniques, we develop two code versions, RI-DCM and sRI-DCM, within a developmental version of Q-Chem\cite{epifanovsky2021software} package.
We take the RI-DCM results as references and focus on accuracy, time cost, and convergence pattern to evaluate our sRI-DCM method.
If not stated otherwise, all calculations in this study use the cc-pVDZ basis set.

Because the sRI technique is stochastic, we average the results of sRI-DCM over ten independent runs with different random seeds.
We quantify the stochastic uncertainty using the standard deviation (S.D.).
If the reference RI-DCM value falls within the S.D. range, indicating the absolute error is smaller than the S.D., the sRI-DCM result is considered accurate.
Other parameters follow Ganoe’s work: the maximum DCM order is 20, and the rescale factor $s$ is 1.1.

All the calculations are carried out in the high performance computing (HPC) center of Westlake University, utilizing an AMD EPYC 7502 (2.5 GHz) node with 64 computational cores.

\subsection*{3.1. Accuracy benckmark}

\begin{figure}[htbp]
    \centering{}
    \subfigure[]{
        \includegraphics[width=0.45\linewidth]{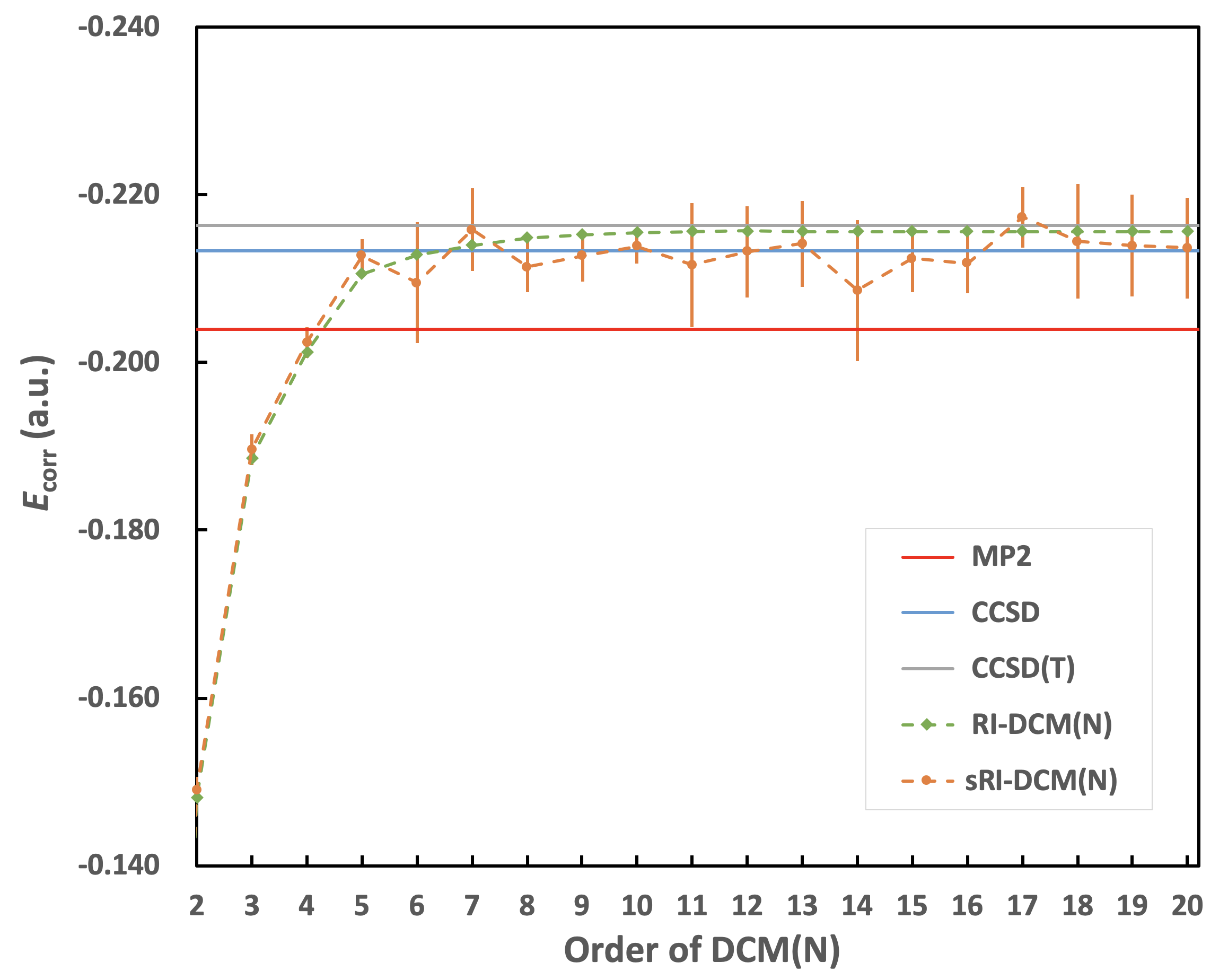}
        \label{fig:benchmark_h2o}
    }
    \hfill
    \subfigure[]{
        \includegraphics[width=0.45\linewidth]{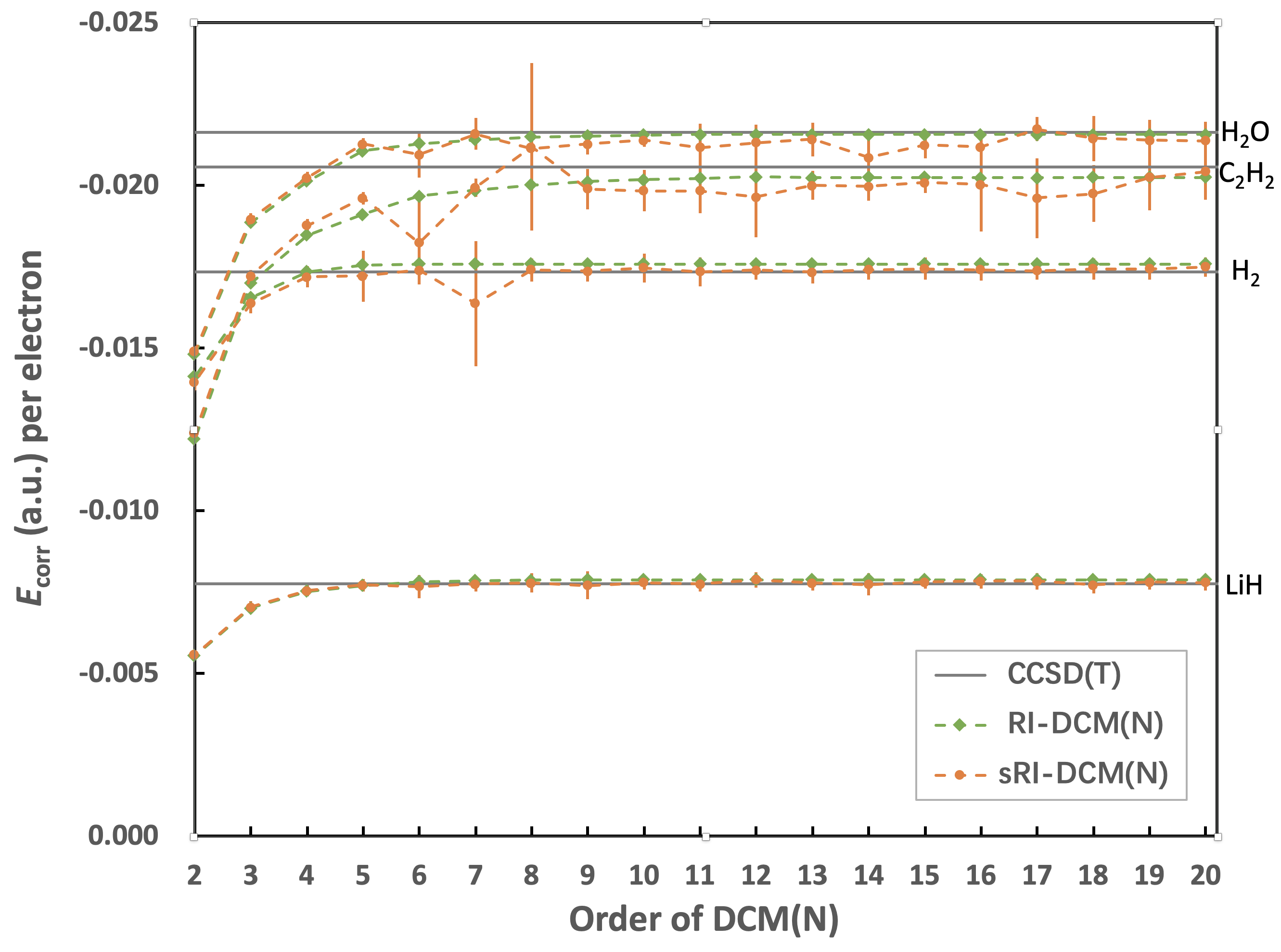} 
        \label{fig:benchmark_small}
    }
    \caption{Error comparison of RI-DCM and sRI-DCM among small molecules. Panel (a) is for $\rm H_2O$ only, and panel (b) adds some other molecules.}
    \label{fig:benchmark}
\end{figure}

\hspace{1em} 
We first benchmark the accuracy of sRI-DCM among a range of small molecules with $N_s=5000$ stochastic orbitals.
As shown in Table \ref{tab:sridcm_benckmark}, the errors of RI-DCM and sRI-DCM are listed for orders $n=5,10,15$ and $20$.
\begin{table}[htbp]
    \caption{Comparison of sRI-DCM and RI-DCM for correlation energy per electron (in $10^{-3}$ a.u.) among small molecules.}
    \small
    \label{tab:sridcm_benckmark}
    \begin{adjustbox}{center}
        \begin{tabular}{ccccccccc}
            \toprule[0.3pt]
            \specialrule{0em}{0.3pt}{0.3pt}
            \midrule
            \multirow{2}*{Molecule} & \multicolumn{2}{c}{DCM(5)} & \multicolumn{2}{c}{DCM(10)} & \multicolumn{2}{c}{DCM(15)} & \multicolumn{2}{c}{DCM(20)} \\
            \specialrule{0em}{1pt}{1pt}
            \cline{2-9}
            \specialrule{0em}{1pt}{1pt}
            & Abs error & S.D. & Abs error & S.D. & Abs error & S.D. & Abs error & S.D. \\
            \midrule
            $\rm H_2$    & 0.3525 & 0.7795 & 0.1329 & 0.4485 & 0.1521 & 0.3338 & 0.1020 & 0.3067 \\
            $\rm H_2O$   & 0.1936 & 0.2159 & 0.1627 & 0.1995 & 0.3266 & 0.3968 & 0.1998 & 0.6000 \\
            LiH          & 0.0152 & 0.1826 & 0.0945 & 0.2206 & 0.0761 & 0.2054 & 0.0851 & 0.2495 \\
            LiF          & 0.0074 & 0.2337 & 0.2489 & 0.2742 & 0.2845 & 0.3938 & 0.4451 & 0.5671 \\
            HF           & 0.1339 & 0.3272 & 0.3086 & 0.7789 & 0.1678 & 0.4896 & 0.2463 & 0.8952 \\
            $\rm NH_3$   & 0.2706 & 0.2723 & 0.1221 & 0.4214 & 0.3788 & 1.4331 & 0.1133 & 0.9891 \\
            $\rm C_2H_2$ & 0.1886 & 0.4921 & 0.3353 & 0.6294 & 0.1732 & 0.3166 & 0.1776 & 0.8656 \\
            \midrule
            \specialrule{0em}{0.3pt}{0.3pt}
            \bottomrule[0.3pt]
        \end{tabular}
    \end{adjustbox}
\end{table}
Using water ($\rm H_2O$) as an example, the results are shown in Figure \ref{fig:benchmark_h2o}.
As the DCM order increases, the RI‑DCM curve smoothly approaches the CCSD(T) reference line by around $n=10$.
The sRI‑DCM curve also converges at a similar order.
All RI‑DCM points lie within the error bars, confirming the validity of the sRI approximation.
In practice, larger $N_s$ values can be used for greater accuracy, and we will discuss this further in Subsection 3.3.
Figure \ref{fig:benchmark_small} includes other molecules, where similar trends are observed.

\subsection*{3.2. Scaling analysis}

\begin{figure}[htbp]
    \centering{}
    \includegraphics[width=8cm]{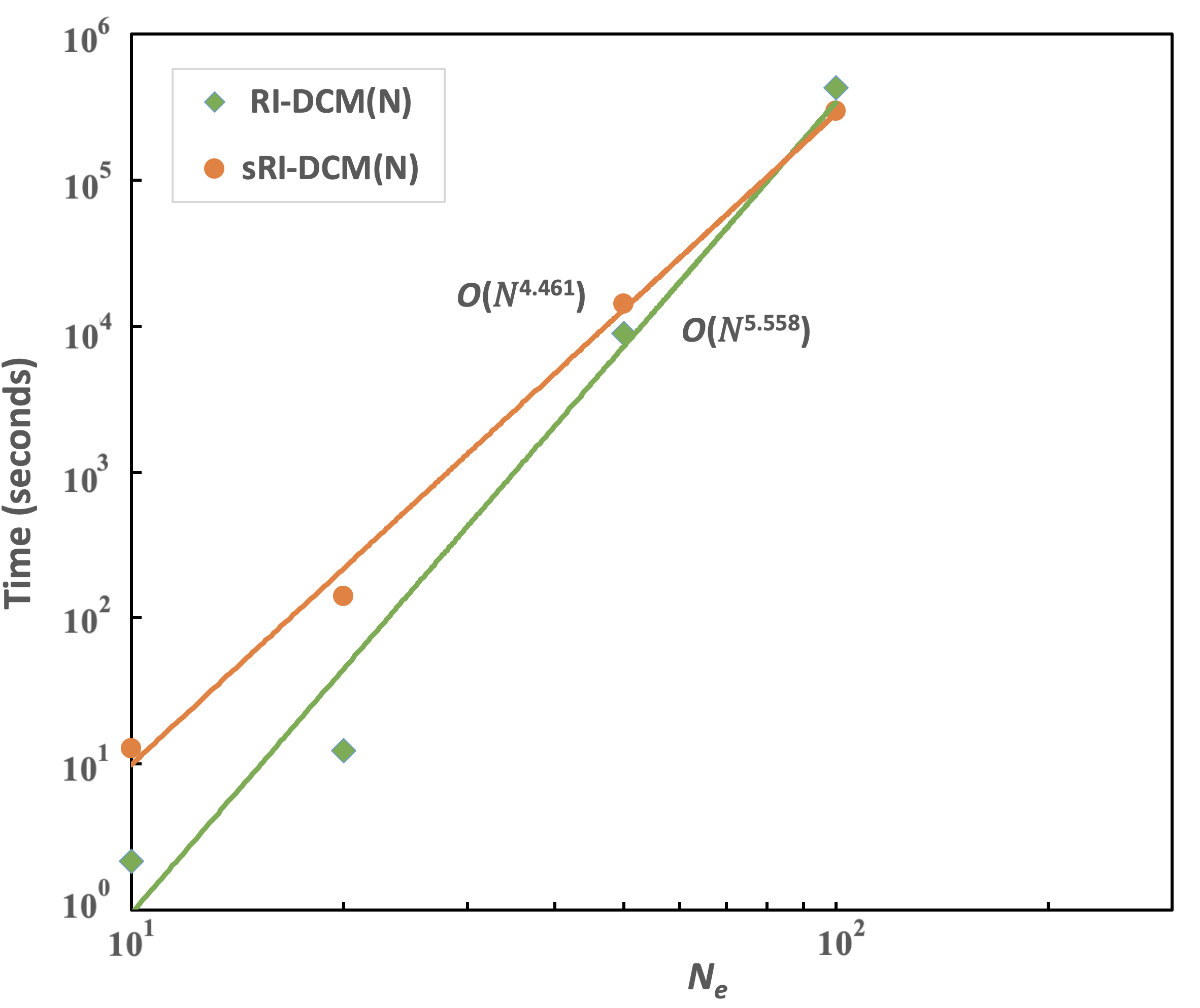} 
    \caption{Comparative scaling of computational cost for sRI-DCM and RI-DCM.}
    \label{fig:scaling}
\end{figure}

\hspace{1em} 
We then assess the computational time cost of our sRI-DCM within a series of hydrogen dimer chains using sto-3g basis set and $N_s = 5000$.
In Figure \ref{fig:scaling}, we presents the fitted curves of time consumption as a function of the chain size.
As shown in the figure, the experimental scaling for RI-DCM is $O(N^{5.58})$, while the scaling of sRI-DCM is $O(N^{4.46})$.
Overall, our sRI-DCM reduces the computational cost by approximately one order of magnitude.
This improved scaling makes our method more efficient than the baseline for systems smaller than the crossover point at around $N_e = 100$.

\subsection*{3.3. Convergence with $N_s$}

\begin{figure}[htbp]
    \centering{}
    \includegraphics[width=8cm]{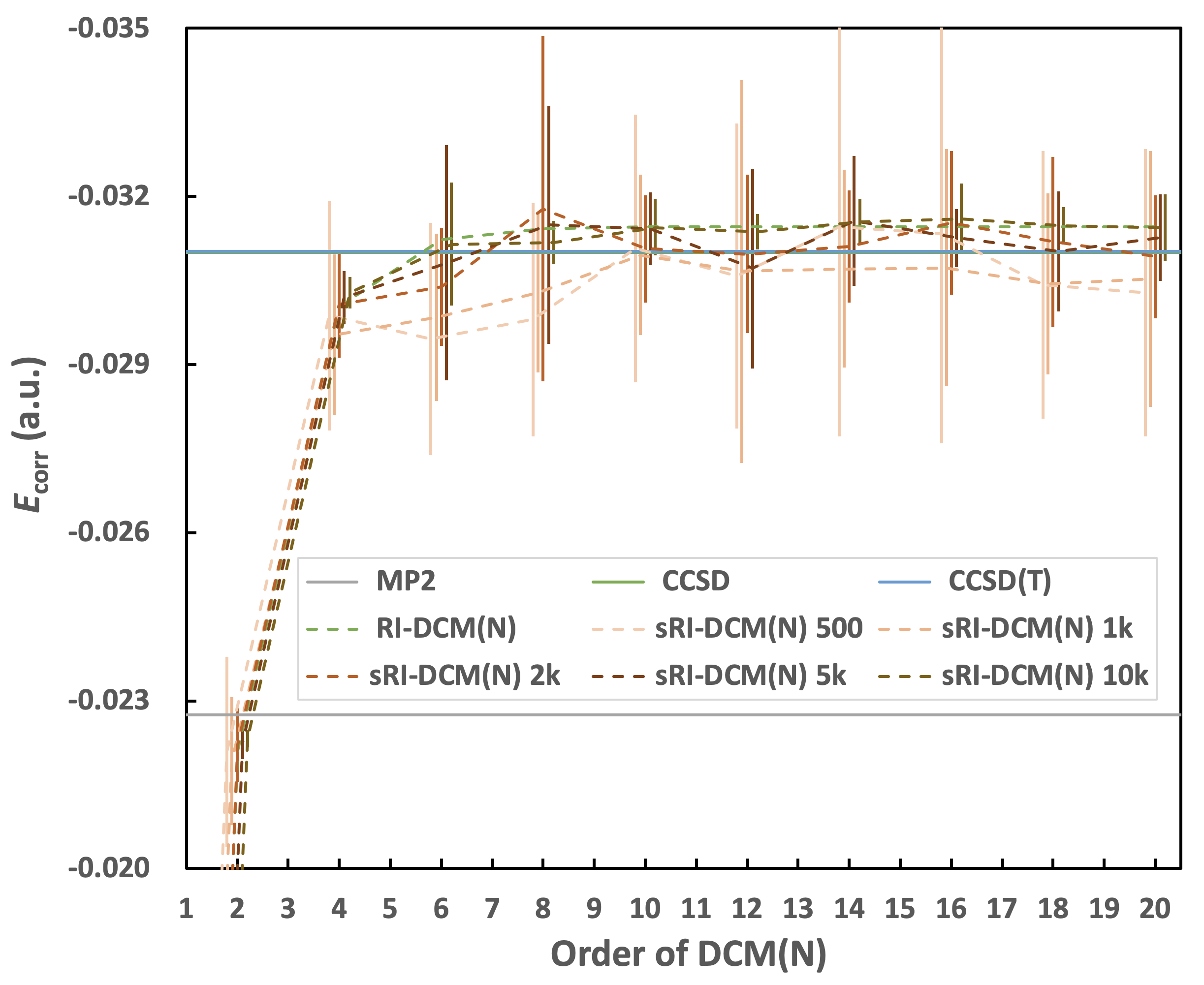} 
    \caption{Correlation energy of LiH with RI-DCM and sRI-DCM using different $N_s$.}
    \label{fig:Ns}
\end{figure}

\hspace{1em} 
In Figure \ref{fig:Ns}, we take LiH as an example to evaluate the convergence performance of sRI-CC2 with different $N_s$ values.
For clarity, the x-axis values are slightly shifted for each $N_s$.
Overall, the sRI-DCM results agree well with RI-DCM across the tested DCM orders.
Notably, larger $N_s$ values lead to smaller error bars (S.D.), which are expected to approach zero. 
This indicates that using more stochastic orbitals helps reduce stochastic noise and improves accuracy.
However, increasing $N_s$ also raises computational cost.
So in practice, we need to choose an appropriate $N_s$ that balances accuracy and efficiency and in our calculations above, we select $N_s = 5000$.

\section*{4. CONCLUSIONS}

\hspace{1em} 
In this work, we present a sRI formulation of DCM to estimate correlation energy.
Our sRI-DCM achieves good agreement with the original DCM while reducing the computational scaling from $O(N^6)$ to approximately $O(N^4.5)$.
It maintains a similar convergence pattern, and its accuracy can be controlled in practice by adjusting the number of stochastic orbitals $N_s$.

The use of sRI in this work extends its application from traditional methods to moment-based calculations.
This demonstrates the flexibility and strong scaling reduction capability of sRI in different situations.
Our future interest lies in incorporating other techniques, such as tensor hypercontraction (THC),\cite{hohenstein2012tensor,hohenstein2022rank,lee2019systematically} to further lower the scaling to $O(N^3)$ and reduce stochastic noise.
Besides, we will use moment expansion to calculate excitation energy.

\section*{AUTHOR DECLARATIONS}
\subsection*{Conflict of Interest}
\hspace{1em} 
The authors have no conflicts to disclose.

\section*{APPENDIX: RESULTS OBTAINED FROM THE CHOLESKY DECOMPOSITION SCHEME}

\hspace{1em} 
As mentioned in Section 2.3, using Cholesky decomposition to decouple the crucial four-index intermediate is also a useful scheme.
However, since its performance in accuracy is inferior than eigendecomposition, we only provide some data here.

\begin{figure}[htbp]
    \centering{}
    \subfigure[]{
        \includegraphics[width=0.45\linewidth]{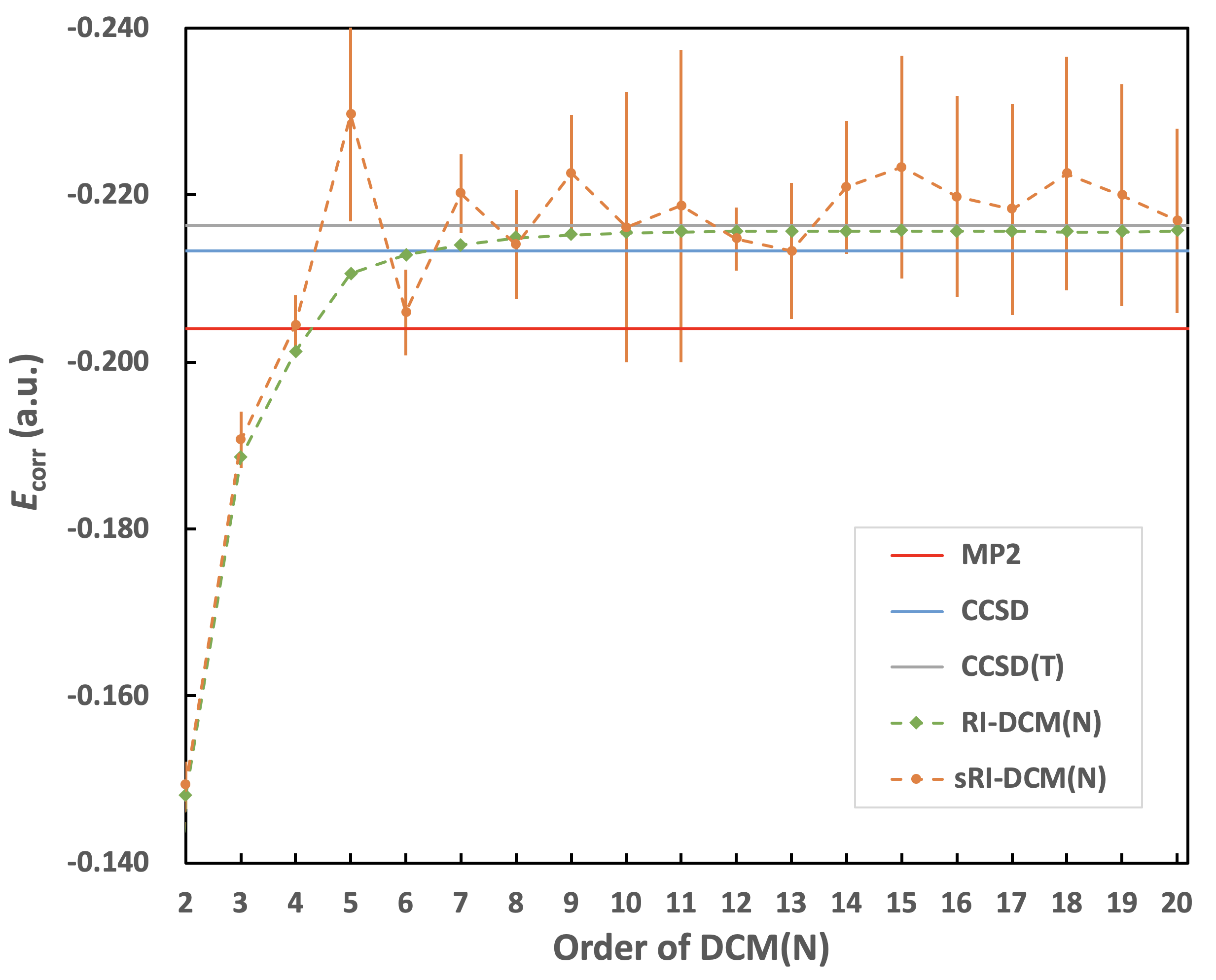}
        \label{fig:benchmark_h2o_cd}
    }
    \hfill
    \subfigure[]{
        \includegraphics[width=0.45\linewidth]{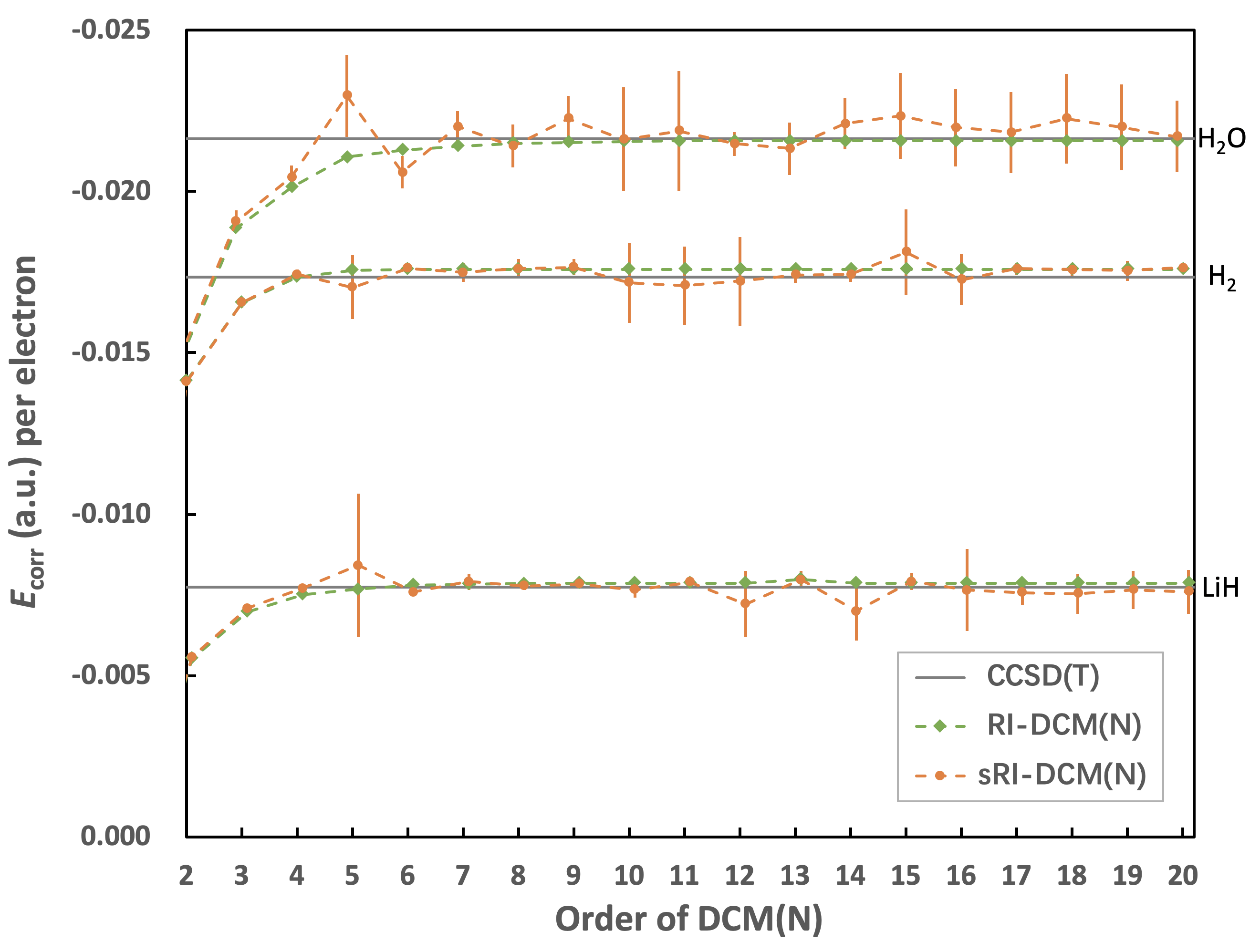} 
        \label{fig:benchmark_small_cd}
    }
    \caption{Error comparison of RI-DCM and sRI-DCM among small molecules. Panel (a) is for $\rm H_2O$ only, and panel (b) adds some other molecules.}
    \label{fig:benchmark_cd}
\end{figure}

In Figure \ref{fig:benchmark_cd}, we plot the correlation energy of some small molecules.
Still, the left subfigure corresponds to $\rm H_2O$ and the right includes additional molecules.
Moreover, the same axes as in Figure \ref{fig:benchmark} are used to ensure a clear comparison.
By comparing Figure \ref{fig:benchmark_h2o_cd} with Figure \ref{fig:benchmark_h2o}, we can observe that our sRI-DCM with Cholesky decomposition yields larger errors.
Nevertheless, its results remain in good agreement with those of RI-DCM, and we therefore regard it as a potentially useful attempt in our theoretical development.

\section*{ACKNOWLEDGEMENTS}

\hspace{1em} 
We acknowledge the high-performance computing (HPC) service from Westlake University.
W.D. thanks the funding from National Natural Science Foundation of China (No. 22361142829) and Zhejiang Provincial Natural Science Foundation (No. XHD24B0301).
We are grateful for helpful discussions from Joonho Lee.

\section*{DATA AVAILABILITY}

\hspace{1em} 
The data that support the findings of this study are available within the article.

\bibliographystyle{IEEEtran} 
\bibliography{ref} 

\end{document}